# A $Sc_2C_2@C_{88}$ cluster based ultra-compact multi-level probabilistic bit for matrix multiplication

Haoran Qi[1,2,3,#], Guohao Xi[4,#], Yuan–Biao Zhou[5,#], Xinrong Liu[1,2,3,#], Yifu Mao[1,2,3,#], Jian Yang[1,2,3], Jun Chen[1,2,3], Kuojuei Hu[1,2,3], Weiwei Gao[2,3], Shuai Zhang[1,2,3], Xiaoqin Gao[1], Jianguo Wan[1], Da–Wei Zhou[6], Junhong An[7], Xuefeng Wang[8], De–Chuan Zhan[6], Minhao Zhang[1,2,3,*], Cong Wang[4,*], Wei ji[4], Yuan–Zhi Tan[5,*], Su–Yuan Xie[5], Fengqi Song[1,2,3*]

[1]National Laboratory of Solid State Microstructures, Collaborative Innovation Center of Advanced Microstructures, School of Physics, Nanjing University, Nanjing 210093, China

[2]Institute of Atomic Scale Manufacturing, Nanjing University, Suzhou 215163, China

[3]Nanjing Institute of Atomic Scale Manufacturing, Nanjing 211800, China

[4]Beijing Key Laboratory of Optoelectronic Functional Materials and Micro-Nano Devices, Key Laboratory of Quantum State Construction and Manipulation(Ministry of Education), Department of Physics, Renmin University of China, Beijing 100872, China

[5]State Key Laboratory of Physical Chemistry of Solid Surfaces, College of Chemistry and Chemical Engineering, Xiamen University, Xiamen 361005, China

[6]National Key Laboratory for Novel Software Technology, School of Artificial Intelligence, Nanjing University, Nanjing 210023, China

[7]Key Laboratory of Quantum Theory and Applications of MoE, Lanzhou Center for Theoretical Physics, and Key Laboratory of Theoretical Physics of Gansu Province, Lanzhou University, Lanzhou 730000, China

[8]Jiangsu Provincial Key Laboratory of Advanced Photonic and Electronic Materials, State Key Laboratory of Spintronics Devices and Technologies, School of Electronic Science and Engineering, Collaborative Innovation Center of Advanced Microstructures, Nanjing University, Nanjing 210093, China

[#]These authors contributed equally: Haoran Qi, Guohao Xi, Yuanbiao Zhou, Xinrong Liu, Yifu Mao

[*]Corresponding authors. Email: M.Z. (zhangminhao@nju.edu.cn), C.W. (wcphys@ruc.edu.cn), Y.-Z. T. (yuanzhi_tan@xmu.edu.cn), and F.S. (songfengqi@nju.edu.cn)

**Abstract**

Information units are progressively approaching the fundamental physical limits of the integration density, including in terms of extremely small sizes, multistates and probabilistic traversal. However, simultaneously encompassing all of these characteristics in a unit remains elusive. Here, via real-time in situ electrical monitoring, we clearly observed stochastic alterations of multiple conductance states in $Sc_2C_2@C_{88}$. The true random bit sequence generated exhibited an autocorrelation function whose confidence interval fell within ±0.02, demonstrating high-quality randomness. The alterations of multiple conductance states are controllable, that is, whose probability distributions could traverse from "0" to "1", enabling us to factorize 551 into its prime factors. Furthermore, we proposed a matrix-chain multiplication scheme and experimentally verified the multiplication of two 4 × 4 state-transition matrices with a small maximum error < 0.05. Combined with theoretical calculations, the stochastic but controllable multistates are probably attributed to the rich energy landscape, which could be stepwise changed by the electric field. Our findings reveal extremely small multi-level probabilistic bit for matrix multiplication, which pave the way for ultracompact intelligent electronic devices.

**Background**

Owing to the proliferation of big-data storage, machine learning and scientific computing applications, advanced information units are currently in greater demand than ever before[1,2]. First, there is an increasing demand for extremely small information units, such as storage or computing based on DNA[3-6], single atoms[7-11], and single molecules[12-17], as they allow a significantly increased data capacity within the same physical space. In addition, a multistate capability represents another critical requirement for advanced information units, as it enables a significant increase in data density by allowing storage of more than one bit of information per cell[18-22].

Furthermore, probabilistic traversal within a single information unit, designated as a probabilistic bit (p-bit), has generated considerable interest in the development of alternative computing paradigms that address uncertainty issues and combinatorial optimization challenges[23,24]. For example, in prime factorization based on magnetic tunnel junctions (tens of nanometers), a low-energy barrier enables stochastic switching of the magnetization between two states, assisted by thermal fluctuations[23]. Other nanosystems, such as magnetic skyrmions and manganite nanowires, also have the potential to implement the p-bit functionality[25,26].

However, demonstrating a unit that possesses all of these characteristics, i.e., an extremely small size, a multistate capability, and probabilistic traversal, is extremely difficult. Previous studies have demonstrated stochastic alterations in the conductance as seen by the scanning tunneling microscopy technique at the atomic level[27-29], but translating these observations into scalable, batch-fabricated devices has been challenging. Recent studies on single-cluster devices of endohedral fullerenes indicate the electrically switchable electret/magnetism, carbon cage protection and possible multistable states[16,17,30], which shed lights on a practical ultra-compact multi-level probabilistic bit.

Here we report the electrically switchable stochastic alterations in the conductance in a single $Sc_2C_2@C_{88}$ cluster device. We investigate the statistical distributions and transition probability matrices of multiple conductance states. On the basis of the ultracompact information unit, we showcase the probabilistic traversal capability in prime factorization of 551 and a high-precision multiplication of two 4 × 4 matrices. Together with the theoretical calculations, we reveal that the rich energy landscape stepwise evolves with the electric field, which is likely to induce stochastic multistates and matrices.

**Single-electron transport of $Sc_2C_2@C_{88}$ cluster**

Here, we selected $Sc_2C_2@C_s$(hept)-$C_{88}$ (< 1 nm) as an ultracompact unit (**Figure 1a**) because the symmetry of the $C_{88}$ cage, the orientation of the $Sc_2C_2$ molecule, and the alignment of the $C_2$ dimer collectively endow it with a rich energy landscape. Here, the energy landscape describes the relationship between the potential energy surface of the $Sc_2C_2@C_{88}$ cluster and its internal configuration of the encapsulated $Sc_2C_2$ unit. Our devices feature a transistor architecture comprising source, drain, and gate electrodes, as shown in the SEM image in Figure 1a. At its core lies an hourglass-shaped nanowire (~50 nm at its narrowest) fabricated via electron-beam lithography (EBL), which is highlighted in pink in **Figure 1a** and magnified in the inset of **Figure 1b**. We then need to create a nanogap within this nanowire to establish a tunneling contact between a $Sc_2C_2@C_{88}$ cluster and the gold electrodes. Here, we used the feedback-controlled electromigration break junction (FCEBJ) method at 1.8 K[16,30]. The source–drain current ($I_{sd}$)–voltage ($V_{sd}$) relationship during the FCEBJ process is shown in Fig. 1b. The red line is the starting line, whereas the blue line is the ending line. The $Sc_2C_2@C_{88}$ solution was coated onto the devices in a glove box before we conducted the FCEBJ procedure. More details regarding the device fabrication are provided in **Supplementary Information 1**.

After the FCEBJ process, we conducted $I_{sd}$–$V_{sd}$ measurements to identify whether the $Sc_2C_2@C_{88}$ transistors were successfully prepared. Unless otherwise noted, the current-voltage characteristics of devices shown were measured at a cryostat temperature of 2 K. For successfully prepared $Sc_2C_2@C_{88}$ transistors, the $I_{sd}$–$V_{sd}$ curve usually reveals a nonlinear property with distinct Coulomb blockade in the low-voltage region (**Figure 1c**). The region could be further manipulated by the gate voltage, indicating that the electric field effectively manipulates the electrochemical potential of $Sc_2C_2@C_{88}$. **Figure 1d** shows the mapping of the differential conductance ($dI/dV$) as a function of the bias voltage ($V_{sd}$) and gate voltage ($V_g$) in a $Sc_2C_2@C_{88}$ transistor (defined as Device 1). The blue regions represent the Coulomb blockade regime, whereas the white and red regions indicate differential conductance peaks. For clarity, the Coulomb edges are marked with black dashed lines. On the basis of the Coulomb diamond pattern in **Figure 1d**, the maximum charging energy of the device could be estimated (exceeding 75 meV), which indicated that the transport signal originates from a $Sc_2C_2@C_{88}$ rather than from a gold nanoparticle formed due to electromigration. **Figure 1e** shows the current–gate voltage relationship extracted from the current mapping corresponding to **Figure 1d**. Significantly, the source–drain current randomly switches among at least two distinct current levels at $V_g = 0$ V, which are marked with red and blue lines, respectively (**Figure 1f**). In this study, the probability of obtaining measurable tunneling currents after the FCEBJ process was approximately 60%. Following FCEBJ, a total of 381 devices exhibited measurable tunneling currents. Among these, Coulomb blockade features were observed in approximately 280 devices, and stochastic switching behavior was demonstrated in 73 devices. Examples of the fabricated stochastic switching elements are presented in **Supplementary Information 2 and Fig. S2**.

**Stochastic alterations of multiple conductance states**

Among all the devices that exhibited stochastic behavior, we take Device 2 as an example to investigate the detailed stochastic alterations of multiple conductance states. The current–time ($I$–$t$) measurements at $V_{sd} = 110$ mV exhibit random switching between three conductance states, as evidenced by the distinct current levels (**Figure 2a**). A statistical histogram of the current distribution is plotted in **Figure 2b**, revealing three discrete states marked in orange, red, and blue. As shown in **Figures 2c, e**, we measured the current–time response at different bias voltages, and the current distributions are plotted in **Figures 2d, f**. The number of states and their distribution significantly change with the bias voltage: the device exhibits four discrete states at 140 mV and only three discrete states at 180 mV.

We also investigated the randomness of the state transitions. We selected two sets of current–time responses at 140 mV and generated a 0–1 random number sequence. The confidence interval for the autocorrelation coefficient (ACF) of the sequence lies within ±0.02 (**Figure 2g**). The ACF could be further decreased by increasing the measurement time (**Figure S3**). Compared with existing works[31-35], our autocorrelation results exhibit high-quality randomness. We also plot the 0–1 random number sequence in code maps (**Figure 2h**), confirming the high-quality true random sequence generated.

**Controllable alterations of multiple conductance states**

Then, we measured the multiple conductance states at more bias voltages in Device 2 and obtained the statistical distribution of the multistates, as shown in **Figure. 3a–f**. As the voltage increases, the proportion of the two upper conductance states gradually decreases, while that of the lower conductance states gradually increases. We marked a comparative broken line based on the current distribution of the device, as shown by the yellow line in **Figure 3g**. The areas above and below the broken line are marked as 1 and 0, respectively. We plotted the probability distribution of the 0–1 states versus the bias voltage and fitted it with a sigmoid function (**Figure 3h**), indicating that the probability can be controlled between "0" and "1". The p-bit function was also observed in other $Sc_2C_2@C_{88}$ devices, as shown in **Supplementary information 4**. Although stochastic conductance-state alterations have previously been observed in fullerenes[27,28], probabilistic traversal has not been achieved.

Controllable alterations of multiple conductance states enable us to realize the probabilistic computing, like the integer factorization problem. The goal of the algorithm is to find the p-bit configuration that minimizes $E$ (see the "Factorization Algorithm" section in the **Methods**). Specifically, we saved the voltage to be input in the form of an array, applied it sequentially to the p-bit unit, and recorded the output of the p-bit unit in an array one by one (See **Supplementary information 4** for details). In this way, it equivalently reproduces the effect of multiple p-bits operating in parallel. The $Sc_2C_2@C_{88}$ device serves as the genuine, physical random number generator (p-bit). Specifically, a Keithley 2450 source meter functions simultaneously as a precision voltage source (DAC) and a current meter (ADC), applying the input bias to the device and directly reading its stochastic conductance state. A custom LabVIEW program processes the real-time, probabilistic output stream from the $Sc_2C_2@C_{88}$ device, calculates the next input voltage based on the cost function, and immediately commands the source meter to apply it. This creates a closed-loop, adaptive physical process where each iteration of the algorithm depends on the instantaneous, unpredictable state of the $Sc_2C_2@C_{88}$ device. **Figure 3i** demonstrate the integer factorization results. For each integer to be factored, we repeated the calculation 20 times, and each calculation converged to the correct result: for 551, we found two peaks at (19, 29) and (29, 19). Lower panel of Fig. 3i display the averaged probability distribution of the p-bit at each factorization termination. For example, for the target vector "1, 0, 1, 1" (decimal 29 = $1 \times 2 + 0 \times 4 + 1 \times 8 + 1 \times 16 + 3$), the corresponding probabilities are 0.53, 0.41, 0.68, and 0.89. Positions where the target is 1 are driven toward higher probabilities, whereas positions where the target is 0 are driven toward lower probabilities, confirming the effectiveness of algorithmic control over the device.

**High-precision matrix multiplication**

In addition, we measured the probabilities of the device switching between different states within a specific time interval. As the probability distribution of the state in the next time interval depends solely on its state in the present time interval (see the **Discussion** section for details) and remains constant over time (**Supplementary information 5 and Fig. S5**), the sequence of state transitions forms a continuous-time Markov chain[36-40]. Through the Markov chain model, we represented the state transitions of the device in the form of a state transition matrix. Exploiting the Markov chain's intrinsic properties, we designed tailor-made physical processes that can be tracked in real time, enabling high-precision matrix multiplication (**Supplementary Information 6**).

At the core of the entire experiment lies a Markovian state switching process based on the multi-state $Sc_2C_2@C_{88}$ device. Experimentally, a Keithley 2450 source-measure unit serves a dual purpose: it supplies a bias voltage to modulate the energy landscape of the cluster, thereby altering the state-transition matrix of the device, while simultaneously reading the real-time current signal to determine the device state. Control and data processing are performed by LabVIEW and MATLAB routines, which periodically adjust the bias voltage of the device, acquire the latest conductance values, determine the corresponding state, and accumulate the statistical results to perform matrix multiplication. In this way, each algorithmic step is directly conditioned on the intrinsic and unpredictable state transition of the cluster, effectively transforming the hardware itself into a high-precision matrix multiplier.

For Device 2, we selected two voltages of 140 mV and 160 mV to demonstrate the matrix multiplication process described in **Supplementary Information 6**. With a time interval of 1 s, we recorded the initial and final states of the device under different voltage configurations, and each pair of initial and final states was recorded as one cycle. **Figures 4a and d** show the state–cycle responses under biases of 140 mV and 160 mV, respectively, with four distinct conductance states marked in each plot. **Figures 4b, e** show the state distribution histograms statistically obtained from the state–cycle relationships in **Figure. 4a, d**. **Figures 4c, f** show the state transition matrices analyzed from the data in **Figure. 4a, d**, using a time interval of 1 s and rounding of matrix elements to two decimal places. Furthermore, by repeatedly switching the voltage between 140 mV and 160 mV, we obtained the state–cycle response shown in **Figure 4g**. The voltage was held for 1 s at each level; thus, the total duration of a single cycle was 2 s. **Figure 4h** shows the state distribution corresponding to **Figure 4g**. By statistically analyzing the transition from the initial state to the final state within each cycle, we obtained the transition matrix shown in **Figure 4i**. The matrix elements match the product of the matrices shown in **Figure 4c, f**. **Figure 4j** presents a comparison between the measured matrix (yellow data) and the results obtained through direct calculation (blue data). The maximum and the average errors among the matrix elements are less than 0.05 and 0.03 (see **Fig. S6-2** in **Supplementary information 6**), respectively, thereby validating the high precision of matrix multiplication [41].

For comparison, memristor arrays are commonly employed in matrix multiplication, in which the conductance weights of each device serve as the matrix elements. These conductance weights are applied to the input signals through Kirchhoff's laws, thereby naturally implementing matrix multiplication[21,42-45]. In $Sc_2C_2@C_{88}$ devices, matrix elements and multiplication can be realized all in one device. In addition to the device structure mentioned in **Figure 1**, we have designed and fabricated a four-nanowire array to further illustrate the small computing units and the potential of high integration density (see **Fig. S1-9–Fig. S1-12** in **Supplementary information 1**). We also demonstrated the probabilistic integer factorization and the state-transition-based matrix multiplication successfully (**Supplementary information 8**).

Our current implementation relies on the intrinsic, fixed state transition matrices of the device at specific bias voltages, which differs from the flexible, user-programmable matrix elements typical of memristor crossbar arrays. A future practical pathway could involve mapping a library of useful matrices ($T_1$, $T_2$, ... $T_n$) accessible under specific bias conditions. A user could then select and sequence these pre-defined matrices from the library to perform desired computations, which establishes a novel paradigm for "programming" computational operations in a single-cluster device by exploiting its rich physical tuning methods. To further unlock the potential of matrix, we carried out a series of simulations. We could construct a Gaussian-type error production function capable of generating 7 distinct error levels based on one 4×4 matrix, and simulate the neuromorphic diffusion process (**Supplementary information 9**). Additionally, using the matrices as the weights, we successfully emulated color recognition and classification (**Supplementary information 10**). A training set consisting of 100 red, 100 green, and 100 blue samples is employed. The training error rapidly decreased to approximately 10 misclassified samples within the first five epochs.

**Discussion**

Real-time in-situ electrical measurements have been shown to be useful for probing latent electronic characteristics in devices[46]. The stochastic switching phenomenon is usually related to the evolution of the electronic states, which can be caused by evolution of the atomic structure when an electrical excitation is injected[27-29,47], geometrical fluctuations of the contact[48-50], or the influence of localized traps[51,52]. In our devices, the discrete states can remain stable for a long time according to the statistical histogram of the current distribution (**Supplementary information 5**). These phenomena indicate that factors related to geometric fluctuations are unlikely. Besides, our transport data showed well-defined Coulomb diamond patterns (Fig. 1**d**) and differential conductance peaks, along with a bias window specific nature of the switching (**Supplementary information 2**). All of these evidences are inconsistent with the behavior of random external charge traps, further supporting an intrinsic origin. Therefore, we initially attributed the random switching between multiple distinct states to cluster structure evolution assisted by an electric field.

To elucidate the atomic configurations and their electric-field-driven switching mechanisms, we performed density functional theory (DFT) calculations on $Sc_2C_2@C_{88}$. By systematically considering the $C_{88}$ cage symmetry, $Sc_2C_2$ molecule orientation and $C_2$ dimer alignment, we systematically generated 522 potential configurations (see the Methods). **Figures 5a and S11-1** present the mapped potential energy surface (PES) for the $Sc_2C_2$ orientation, revealing five distinct stable configurations (visualized in **Figures 5d and S11-2**) that are at least 30.4 meV more stable than other configurations, with the most stable structure being consistent with previous literature[17,53]. These states exhibit relative energy differences spanning 0.2–99.3 meV (**SI Table S2**), indicating potential electric field tunability. All the configurations share a common structural feature: one Sc atom persistently occupies a C–C bridge site on the carbon cage. Differentiation between groups arises from distinct adsorption sites of the second Sc atom, whereas intragroup variations originate from rotational reorientations of the $C_2$ dimer.

We further performed climbing image nudged elastic band (CI–NEB) calculations to determine the transition pathways and energy barriers between all pairwise combinations of the five stable states (see **SI Table S3**). All five stable states and their transition states exhibit distinct orientations of moderate electric dipole moments (> 0.39 e·Å, blue arrows in **Figure 5d**; **SI Table S4**). This significant dipole variation enables dual electric field control mechanisms: (1) modulating relative state energies and (2) tuning interstate energy barriers. For example, a (0, 0, 500) mV/Å field reduces the state 3 ↔ 1 barrier to zero while preserving the energy degeneracy (**Figure 5b**). A (100, 0, 0) mV/Å field inverts the stability hierarchy between states 2 and 1 while maintaining a moderate barrier of 38 meV (**Figure 5c**). Through Boltzmann statistics, we established a field-controllable probability matrix (calculation details in the Methods). The applied electric field strength governs both the effective size of the matrix, which is determined by the number of accessible states through selective barrier elimination, and the individual transition probabilities via continuous barrier height modulation. As shown in **Figure 5e**, with increasing strength of the electric field along the *y*-direction, the transition probability matrix undergoes simultaneous changes in the probability magnitudes while progressively reducing in size from 5 × 5 to 2 × 2. This allows us to access a set of different, pre-characterized transition matrices from the same device, which also explains why stochastic switching in $Sc_2C_2@C_{88}$ devices is only prominent within a narrow bias window: at low voltages the barrier separating configurations cannot be surmounted, whereas at excessively high voltages the energy landscape is so strongly distorted that the metastable wells vanish altogether. Therefore, the observed stochastic switching is fundamentally the electric-field-assisted rearrangement of the cluster configuration (Video 1 and 2) across its complex energy landscape.

Under this physical mechanism, the state of the device is determined solely by the current positions of the atoms, and any relaxation information from the past cannot be retained. This means that the state transitions of the device within a certain time interval depend only on its state in the previous time interval and are independent of earlier states. Therefore, we can analyze the results using a Markov model and compare them with the experimental results. We also measured the temperature dependence of the stochastic switching (**Supplementary information 12**). The results indicate that temperature does not significantly affect the time constant of the device. This finding suggests that the state transitions in the device are not dominantly caused by thermally activated processes induced by temperature but rather originate from the electric field (also shown in **Supplementary information 13**).

Further, we fabricated devices with $C_{88}$ deposited following the same procedure. Among all tests, 15 devices exhibited Coulomb blockade (**Supplementary Information 14**). In these devices, the characteristic multi-level, voltage-controllable random switching observed in $Sc_2C_2@C_{88}$ was not detected. This further supports that the occurrence of random switching in our system can be primarily attributed to the presence of the encapsulated cluster inside the fullerene cage. Therefore, the intrinsic configurational dynamics of the $Sc_2C_2@C_{88}$ cluster, underpinned by its rich and electrically controllable energy landscape, can be harnessed to realize an ultra-compact, multi-level probabilistic unit. Considering the limitations of current fabrication technique, more deterministic assembly methods would facilitate scalable high-temperature integration, advancing the translation of this material-level discovery into a viable technology for probabilistic and neuromorphic computing.

**Conclusions**

In summary, we investigated the multiple conductance states, probabilistic traversal and state transition matrix in a $Sc_2C_2@C_{88}$ unit. The multiple conductance states switch randomly but controllably, and the confidence interval of the ACF of the random sequence is less than 0.02. On the basis of these characteristics, we demonstrated many potentialities of $Sc_2C_2@C_{88}$, including prime factorization and matrix multiplication. For matrix multiplication, the maximum error is small: less than

0.05. Combined with theoretical calculations, we uncovered the rich energy landscape in the ultracompact unit, which paves the way for ultracompact intelligence all in one.

## Methods

### The fabrication of $Sc_2C_2@C_{88}$ cluster device

See **Supplementary information 1** for details, including substrate preparation and electrode/nanowire patterning, synthesis and characterization of the $Sc_2C_2@C_{88}$ cluster, formation of a single-cluster junction via feedback-controlled electromigration, and etc. It is crucial to clarify here that the nanowire, which forms the core of the device, is patterned using electron-beam lithography (EBL). This nanowire has a nominal minimum width of 50 nm and a length of approximately 400 nm.

### The experimental setup

All transport measurements were performed in a cryogen-free high-magnetic-field cryogenic system (Cryogenic Ltd.) with a base temperature of 1.8 K to ensure electrode stability. The device was wire-bonded to a sample stage, and all electrical connections were made via cryogenic wiring to the room-temperature instrumentation. The Keithley 2450 source meters provided the necessary microvolt precision and current sensitivity for single-cluster transport measurements. The entire measurement sequence—including voltage output, data acquisition, and the real-time control logic—was orchestrated by the LabVIEW program via GPIB communication. Therefore, there is a direct and causal connection between the measured stochastic properties of the $Sc_2C_2@C_{88}$ device and the computing results presented.

### Factorization algorithm

Using a Keithley 2450 as a DAC, we applied inputs to the device and directed the outputs through a circuit board into a computer. The entire process was controlled by a

LabVIEW program, which implemented an algorithm to minimize the cost function $E$. In the experiment, the initial probability of the p-bit was set to 0.5. We saved the voltage that was to be input into the device in the form of an array, sequentially applied it to the p-bit units, and recorded the output of the p-bit units in an array one by one. We assigned the output results to the corresponding bit positions of the two factors and carried out further calculations. The $i$-th bit position could be driven by voltage input $V_i$. The adjustment value for $V_i$ after each learning iteration was calculated as follows:

$$\Delta V_i = -\partial E/\partial m_i$$

where $m_i$ is the output of the $i$-th bit position, expressed in binary form.

Starting from the form of the cost function $E$, we assumed that the number $F$ to be factored could be expressed as the product of two odd numbers $X$ and $Y$, both greater than or equal to 3 (because even numbers can be quickly preprocessed and the factor 1 is trivial). This was achieved via the following relations:

$$E = (XY - F)^2$$

$$X = 3 + \sum_{i=1}^{n} 2^{i-1} m_i$$

$$Y = 3 + \sum_{j=1}^{n} 2^{j-1} m_j$$

Throughout the learning process, we imposed probability constraints, requiring the probability range of the p-bit to be [0.1, 0.9]. This ensured that even when the device entered a local minimum, it could still escape after a few learning iterations. When the learning process resulted in $E = 0$, the loop was exited, and the result was output.

**Theoretical calculations**

Our DFT calculations were performed using the generalized gradient approximation and the projector augmented wave method[54,55] as implemented in the Vienna Ab initio Simulation Package[56]. The Perdew–Burke–Ernzerhof (PBE) function[57] with D3 dispersion correction[58] was employed. The kinetic energy cutoff was set to 400 eV. A 25 × 25 × 25 Å³ supercell was used to model the isolated $Sc_2C_2@C_{88}$ molecule, which ensured a separation of at least 17 Å between the molecule and its images (see **Fig. S11-3** in **Supplementary information 11**). The Γ point was used for sampling the first Brillouin zone in all calculations. All the atoms were allowed to relax until the residual force on each atom was less than 0.01 eVÅ⁻¹.

Transition pathways and energy barriers were revealed by the CI–NEB method[59,60], which locates the exact saddle point of a reaction pathway. The electric dipole moments of the $C_{88}$ molecule system were calculated on the basis of the classical definition:

$$P = \frac{1}{V}\left(-e\sum_j Z_j u_j\right) + \int r\rho(r)dr$$

where $e$ is the electron charge, $V$ is the cell volume, $Z_j$ and $u_j$ are the atomic number and position of atom $j$, respectively, and $\rho(r)$ is the electronic charge density at location $r$ in real space. Dipole correction was considered in all calculations to correct the error introduced by the periodic boundary condition and balance the vacuum level differences on the different sides of the polarized molecules[61,62].

The interstate transition probability model based on the Boltzmann distribution[63,64] can be expressed as follows: For any given microscopic state $i$ in the system, the transition probability to a target state $j$ is given by:

$$P_{i \to j} = \frac{e^{-\Delta E_{ij}/(k_B T)}}{Z_i}$$

$$Z_i = \sum e^{-\frac{\Delta E_{ij}}{k_B T}}$$

where $\Delta E_{ij}$ represents the energy barrier per atom to overcome for the transition from state $i$ to state $j$, $k_B$ is the Boltzmann constant, $T$ is the thermodynamic temperature (2 K) of the system, and $Z_i$ is the partition function for probability normalization.

**Screening of stable configurations of $Sc_2C_2@C_{88}$**

We considered inequivalent adsorption sites for the two Sc atoms of the $Sc_2C_2$ molecule inside the $C_s$(hept)-$C_{88}$ cage, including hollow sites above the $C_5$/$C_6$/$C_7$ rings, C–C bonds, and top sites above the C atoms (totaling 145 types of sites). On the basis of the Sc–C bond length (~2.1 Å) and C–C bond length (~1.2 Å) within the $Sc_2C_2$ molecule, along with the orientation of $C_2$, we constrained the Sc–Sc distance to 3.6–4.7 Å and applied mirror symmetry, ultimately identifying 370 candidate configurations. After identifying the relatively stable $Sc_2C_2$ adsorption configurations, we further considered the spatial orientation of the central $C_2$ dimer bridge within the $Sc_2C_2$ molecule for the five most stable configurations, which led to the calculation of an additional 152 configurations. Following structural optimization and total energy comparison via first-principles calculations, we discarded configurations with significantly higher energies or those unable to maintain structural stability. This screening yielded the five stable configurations mentioned in the main text (Fig. 5). On the basis of the calculated energies of these structures, we constructed a PES for $Sc_2C_2@C_{88}$ (Figs. 5a and S10-1). Linear interpolation was employed to depict the energies between the points on the PES corresponding to the calculated configurations.

**Supplementary information**

This file contains Supplementary information sections 1–15, including Supplementary Figures 1–15 and Tables 1–4, which provide additional information on fabrication, electrical measurements and DFT calculations supporting the main text.

**Data availability**

The authors declare that the main data supporting the findings of this study are available within this article and its Supplementary information. Source data are provided with this paper. Extra data are available from the corresponding authors upon request.

**Acknowledgments**

We acknowledge the financial support of the National Natural Science Foundation of China (Grant Nos. 92577205, 92161201, 12025404, T2221003, 12422410, 92580203, 12474272, 12374043, T2394473, T2394470, 62274085, 92477205 and 52461160327), the National Key R&D Program of China (Nos. 2022YFA1402404 and 2023YFA1406500), the Fundamental and Interdisciplinary Disciplines Breakthrough Plan of the Ministry of Education of China (JYB2025XDXM411), the Natural Science Foundation of Jiangsu Province (Nos. BK20243013, BK20233001, and BK20240166), Young Elite Scientists Sponsorship Program of the Beijing High Innovation Plan, the Fundamental Research Funds for the Central Universities, and the Research Funds of Renmin University of China (Grants Nos. 22XNKJ30 and 24XNKJ17). All calculations for this study were performed at the Physics Laboratory of High-Performance Computing (PLHPC), and the Public Computing Cloud (PCC) of Renmin University of China.

## Contributions

M.Z., Y.–Z.T. and F.S. conceived the idea for the paper. Y.–B.Z., Y.–Z.T. and S.–Y.X. synthesized and characterized the endohedral metallofullerene. H.Q., X.L., Y.M., J.Y. and J.C. fabricated the transistors. H.Q., X.L. and Y.M. constructed the measurements. H.Q. designed the circuit. G.X., C.W. and W.J. conducted the theoretical calculations. H.Q., G.X., Y.–B. Z, X.L., K.H., W.G., S.Z., X.G., J.W., D.–W.Z., J.A., X.W., D.–C.Z., M.Z., C.W., W.J., Y.–Z.T., S.–Y.X. and F.S. analyzed and discussed the data and wrote the paper. F.S. and Y.–Z.T. supervised the project.

## Competing interests

The authors declare that they have no competing interests.

## References

1   Markov, I. L. Limits on fundamental limits to computation. *Nature* **512**, 147-154, (2014).
2   Wang, H. *et al.* Scientific discovery in the age of artificial intelligence. *Nature* **620**, 47-60, (2023).
3   Adleman, L. M. Molecular Computation of Solutions to Combinatorial Problems. *Science* **266**, 1021-1024, (1994).
4   Cherry, K. M. & Qian, L. Scaling up molecular pattern recognition with DNA-based winner-take-all neural networks. *Nature* **559**, 370-376, (2018).
5   Woods, D. *et al.* Diverse and robust molecular algorithms using reprogrammable DNA self-assembly. *Nature* **567**, 366-372, (2019).
6   Lv, H. *et al.* DNA-based programmable gate arrays for general-purpose DNA computing. *Nature* **622**, 292-300, (2023).
7   Schirm, C. *et al.* A current-driven single-atom memory. *Nature Nanotechnology* **8**, 645-648, (2013).
8   Donati, F. *et al.* Magnetic remanence in single atoms. *Science* **352**, 318-321, (2016).
9   Kalff, F. E. *et al.* A kilobyte rewritable atomic memory. *Nature Nanotechnology* **11**, 926-929, (2016).
10  Natterer, F. D. *et al.* Reading and writing single-atom magnets. *Nature* **543**, 226-228, (2017).
11  Hus, S. M. *et al.* Observation of single-defect memristor in an $MoS_2$ atomic sheet. *Nature Nanotechnology* **16**, 58-62, (2021).
12  Vincent, R., Klyatskaya, S., Ruben, M., Wernsdorfer, W. & Balestro, F. Electronic read-out of a single nuclear spin using a molecular spin transistor. *Nature* **488**, 357-360, (2012).
13  Thiele, S. *et al.* Electrically driven nuclear spin resonance in single-molecule magnets. *Science* **344**, 1135-1138, (2014).

| | |
|---|---|
| 14 | Goodwin, C. A. P., Ortu, F., Reta, D., Chilton, N. F. & Mills, D. P. Molecular magnetic hysteresis at 60 kelvin in dysprosocenium. *Nature* **548**, 439-442, (2017). |
| 15 | Kato, C. *et al.* Giant Hysteretic Single-Molecule Electric Polarisation Switching above Room Temperature. *Angewandte Chemie International Edition* **57**, 13429-13432, (2018). |
| 16 | Zhang, K. *et al.* A Gd@$C_{82}$ single-molecule electret. *Nature Nanotechnology* **15**, 1019-1024, (2020). |
| 17 | Li, J. *et al.* Room-temperature logic-in-memory operations in single-metallofullerene devices. *Nature Materials* **21**, 917-923, (2022). |
| 18 | Athmanathan, A., Stanisavljevic, M., Papandreou, N., Pozidis, H. & Eleftheriou, E. Multilevel-Cell Phase-Change Memory: A Viable Technology. *IEEE Journal on Emerging and Selected Topics in Circuits and Systems* **6**, 87-100, (2016). |
| 19 | Goswami, S. *et al.* Decision trees within a molecular memristor. *Nature* **597**, 51-56, (2021). |
| 20 | Rao, M. *et al.* Thousands of conductance levels in memristors integrated on CMOS. *Nature* **615**, 823-829, (2023). |
| 21 | Sharma, D. *et al.* Linear symmetric self-selecting 14-bit kinetic molecular memristors. *Nature* **633**, 560-566, (2024). |
| 22 | Correll, J. M. *et al.* An 8-bit 20.7 TOPS/W Multilevel Cell ReRAM Macro With ADC-Assisted Bit-Serial Processing. *IEEE Journal of Solid-State Circuits*, 1-14, (2025). |
| 23 | Borders, W. A. *et al.* Integer factorization using stochastic magnetic tunnel junctions. *Nature* **573**, 390-393, (2019). |
| 24 | Roques-Carmes, C. *et al.* Biasing the quantum vacuum to control macroscopic probability distributions. *Science* **381**, 205-209, (2023). |
| 25 | Wang, K. *et al.* Single skyrmion true random number generator using local dynamics and interaction between skyrmions. *Nature Communications* **13**, 722, (2022). |
| 26 | Wang, Y. *et al.* Superior probabilistic computing using operationally stable probabilistic-bit constructed by a manganite nanowire. *National Science Review* **12**, nwae338, (2025). |
| 27 | Chandler, H. J., Stefanou, M., Campbell, E. E. B. & Schaub, R. Li@$C_{60}$ as a multi-state molecular switch. *Nature Communications* **10**, 2283, (2019). |
| 28 | Huang, T. *et al.* A Molecular Switch Based on Current-Driven Rotation of an Encapsulated Cluster within a Fullerene Cage. *Nano Letters* **11**, 5327-5332, (2011). |
| 29 | Auwärter, W. *et al.* A surface-anchored molecular four-level conductance switch based on single proton transfer. *Nature Nanotechnology* **7**, 41-46, (2012). |
| 30 | Wang, F. *et al.* Electrically controlled nonvolatile switching of single-atom magnetism in a Dy@$C_{84}$ single-molecule transistor. *Nature Communications* **15**, 2450, (2024). |
| 31 | Yuan, X. H. *et al.* Arbitrary Modulation of Average Dwell Time in Discrete-Time Markov Chains Based on Tunneling Magnetoresistance Effect. *IEEE Electron Device Letters* **45**, 1349-1352, (2024). |
| 32 | Larimian, S., Mahmoodi, M. R. & Strukov, D. B. Lightweight Integrated Design of PUF and TRNG Security Primitives Based on eFlash Memory in 55-nm CMOS. *IEEE Transactions on Electron Devices* **67**, 1586-1592, (2020). |
| 33 | Ding, Q. T. *et al.* Unified 0.75pJ/Bit TRNG and Attack Resilient 2F/Bit PUF for Robust Hardware Security Solutions with 4-layer Stacking 3D NbO Threshold Switching Array. *2021 IEEE International Electron Devices Meeting (IEDM)*, (2021). |
| 34 | Shen, B. *et al.* Harnessing microcomb-based parallel chaos for random number generation and |

optical decision making. *Nature Communications* **14**, 4590, (2023).

35   Ravichandran, H. *et al.* A stochastic encoder using point defects in two-dimensional materials. *Nature Communications* **15**, 10562, (2024).

36   Preston, R. J., Kershaw, V. F. & Kosov, D. S. Current-induced atomic motion, structural instabilities, and negative temperatures on molecule-electrode interfaces in electronic junctions. *Physical Review B* **101**, 155415 (2020).

37   Rudge, S. L. & Kosov, D. S. Distribution of waiting times between electron cotunneling events. *Physical Review B* **98**, 245402 (2018).

38   Rudge, S. L. & Kosov, D. S. Fluctuating-time and full counting statistics for quantum transport in a system with internal telegraphic noise. *Physical Review B* **100**, 235430 (2019).

39   Niazi, S. *et al.* Training deep Boltzmann networks with sparse Ising machines. *Nature Electronics* **7**, 610-619, (2024).

40   Tian, H. *et al.* A hardware Markov chain algorithm realized in a single device for machine learning. *Nature Communications* **9**, 4305, (2018).

41   Sun, Z. *et al.* Solving matrix equations in one step with cross-point resistive arrays. *Proceedings of the National Academy of Sciences* **116**, 4123-4128, (2019).

42   Duan, X. G. *et al.* Memristor-Based Neuromorphic Chips. *Advanced Materials* **36**, 2310704 (2024).

43   Cai, F. X. *et al.* A fully integrated reprogrammable memristor-CMOS system for efficient multiply-accumulate operations. *Nature Electronics* **2**, 290-299, (2019).

44   Jeong, H. *et al.* Self-supervised video processing with self-calibration on an analogue computing platform based on a selector-less memristor array. *Nature Electronics* **8**, 168-178, (2025).

45   Zhang, W. B. *et al.* Edge learning using a fully integrated neuro-inspired memristor chip. *Science* **381**, 1205-1211, (2023).

46   Kim, Y. & Song, H. Noise spectroscopy of molecular electronic junctions. *Applied Physics Reviews* **8**, 011303, (2021).

47   Guo, Y. *et al.* Emergent complexity of quantum rotation tunneling. *Science Advances* **11**, eads0503, (2025).

48   Dulić, D. *et al.* Controlled Stability of Molecular Junctions. *Angewandte Chemie International Edition* **48**, 8273-8276, (2009).

49   Kihira, Y., Shimada, T., Matsuo, Y., Nakamura, E. & Hasegawa, T. Random Telegraphic Conductance Fluctuation at Au−Pentacene−Au Nanojunctions. *Nano Letters* **9**, 1442-1446, (2009).

50   Park, Y. *et al.* Atomic-precision control of plasmon-induced single-molecule switching in a metal–semiconductor nanojunction. *Nature Communications* **15**, 6709, (2024).

51   Kim, Y., Song, H., Kim, D., Lee, T. & Jeong, H. Noise Characteristics of Charge Tunneling via Localized States in Metal−Molecule−Metal Junctions. *ACS Nano* **4**, 4426-4430, (2010).

52   Arielly, R., Vadai, M., Kardash, D., Noy, G. & Selzer, Y. Real-Time Detection of Redox Events in Molecular Junctions. *Journal of the American Chemical Society* **136**, 2674-2680, (2014).

53   Chen, C.-H. *et al.* Zigzag $Sc_2C_2$ Carbide Cluster inside a [88]Fullerene Cage with One Heptagon, $Sc_2C_2@C_s$(hept)-$C_{88}$: A Kinetically Trapped Fullerene Formed by $C_2$ Insertion? *Journal of the American Chemical Society* **138**, 13030-13037, (2016).

54   Blöchl, P. E. Projector augmented-wave method. *Physical Review B* **50**, 17953-17979, (1994).

55  Kresse, G. & Joubert, D. From ultrasoft pseudopotentials to the projector augmented-wave method. *Physical Review B* **59**, 1758-1775, (1999).

56  Kresse, G. & Furthmüller, J. Efficient iterative schemes for ab initio total-energy calculations using a plane-wave basis set. *Physical Review B* **54**, 11169-11186, (1996).

57  Perdew, J. P., Burke, K. & Ernzerhof, M. Generalized Gradient Approximation Made Simple. *Physical Review Letters* **77**, 3865-3868, (1996).

58  Grimme, S., Antony, J., Ehrlich, S. & Krieg, H. A consistent and accurate ab initio parametrization of density functional dispersion correction (DFT-D) for the 94 elements H-Pu. *The Journal of Chemical Physics* **132**, 154104, (2010).

59  Hong, J. *et al.* Exploring atomic defects in molybdenum disulphide monolayers. *Nature Communications* **6**, 6293, (2015).

60  Qiao, J., Kong, X., Hu, Z.-X., Yang, F. & Ji, W. High-mobility transport anisotropy and linear dichroism in few-layer black phosphorus. *Nature Communications* **5**, 4475, (2014).

61  Qiao, J. *et al.* Few-layer Tellurium: one-dimensional-like layered elementary semiconductor with striking physical properties. *Science Bulletin* **63**, 159-168, (2018).

62  Zhao, Y. *et al.* Extraordinarily Strong Interlayer Interaction in 2D Layered $PtS_2$. *Advanced Materials* **28**, 2399-2407, (2016).

63  Boltzmann, L. Studies on the Balance of Living Force between Moving Material Points. *Wiener Berichte* **58**, 517-560 (1868).

64  Zierenberg, J., Schierz, P. & Janke, W. Canonical free-energy barrier of particle and polymer cluster formation. *Nature Communications* **8**, 14546, (2017).

**Figures**

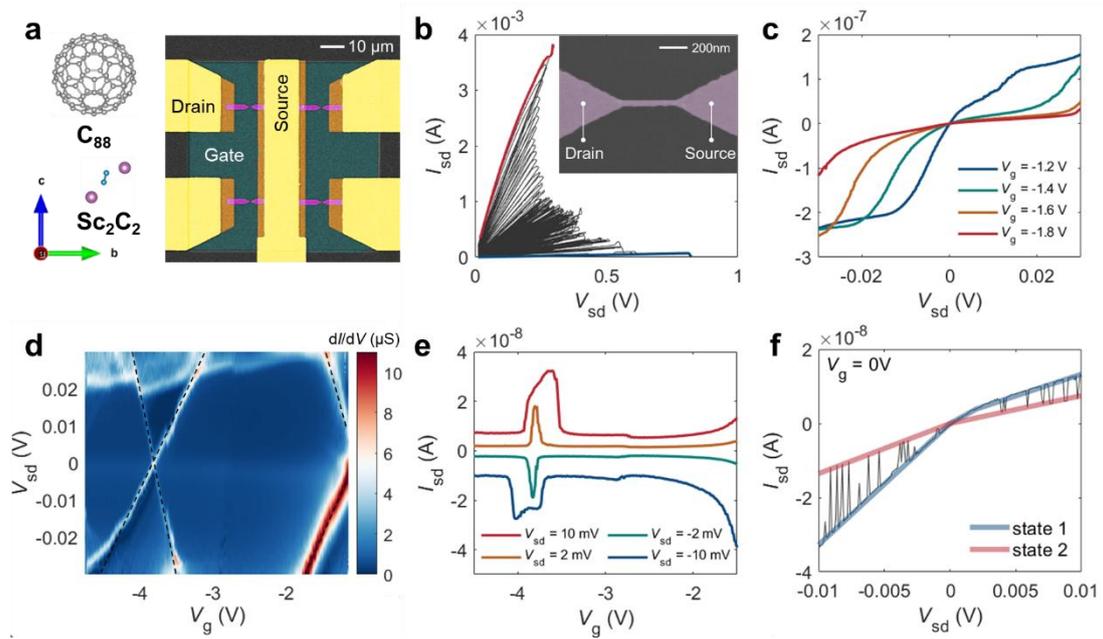

**Fig. 1: The single-electron transport of Sc$_2$C$_2$@C$_{88}$ cluster**

**a**. Schematic diagram and electron micrograph of the Sc$_2$C$_2$@C$_{88}$ device. **b**. Source–drain current ($I_{sd}$)–voltage ($V_{sd}$) relationship during the FCEBJ process. The red line is the starting line, whereas the blue line is the ending line. The illustration shows a scanning electron microscopy image of the hourglass-shaped nanowire before the FCEBJ process. **c**. Coulomb blockade phenomenon at various $V_g$ values in a Sc$_2$C$_2$@C$_{88}$ transistor (Device 1). **d**. Mapping of the differential conductance (d$I$/d$V$) as a function of $V_{sd}$ and $V_g$. **e**. Current–gate voltage relationship extracted from the current mapping corresponding to d. **f**. $I_{sd}$–$V_{sd}$ curve at $V_g$ = 0 V. According to the red and blue markings, Device 1 exhibits random switching behavior.

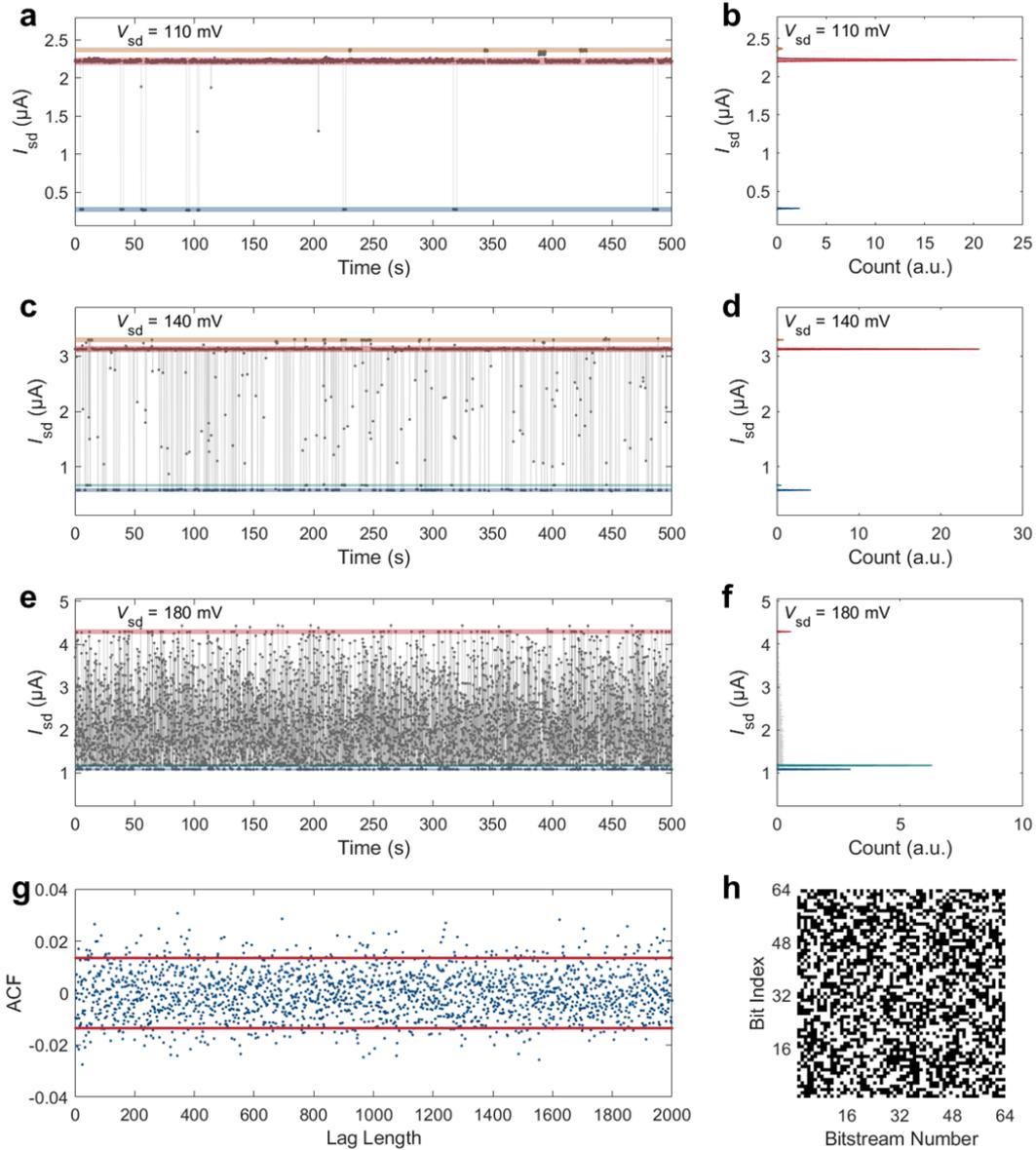

**Fig. 2: Stochastic alterations of multiple conductance states**

**a, c, e.** Random switching between different conductance states at $V_{sd}$ = 110 mV, 140 mV, and 180 mV (Device 2), as shown in the current-time ($I$–$t$) response. The total recording time for each dataset is 500 seconds. **b, d, f.** Current distributions at the corresponding voltages revealing three, four, and three discrete states, respectively. **g.** Autocorrelation functions generated from two sets of current-time data using Device 2 at a bias voltage of 140 mV. The autocorrelation coefficient is less than 0.02 (95% confidence interval). **h.** A 64 × 64 code map plotted using the data from g.

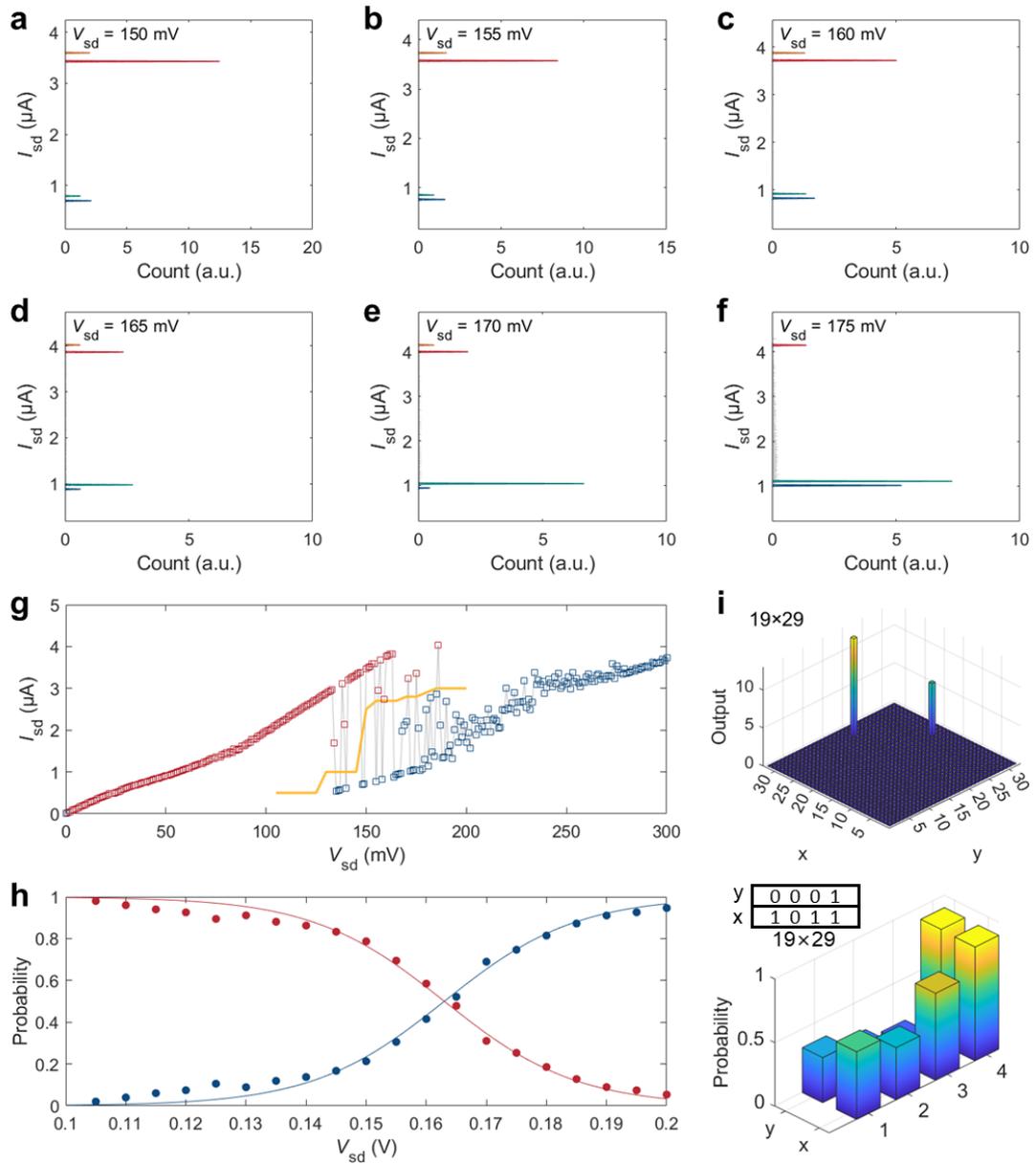

**Fig. 3: Controllable alterations of multiple conductance states**

**a-f.** The regulation of voltage on the cluster conductance states, showcasing six different voltages (Device 2). As the voltage increases gradually, the probability of the high conductance states decreases, while the probability of the low conductance states increases. **g.** The current-voltage response of the device and the division of conductance states. As shown in the figure, the yellow dashed line divides the device into two regions, high and low, corresponding to 0 and 1, respectively. **h.** Voltage control of probability, demonstrating the gradual transition of device probability with voltage variation. **i.** Integer factorization results. For each integer to be factored, we performed 20 repeated calculations, all of which converged to the correct results. The figure illustrates the factorization of 551 using 8 bit positions. The data plot below shows the average probability distributions of the device at the termination of each factorization.

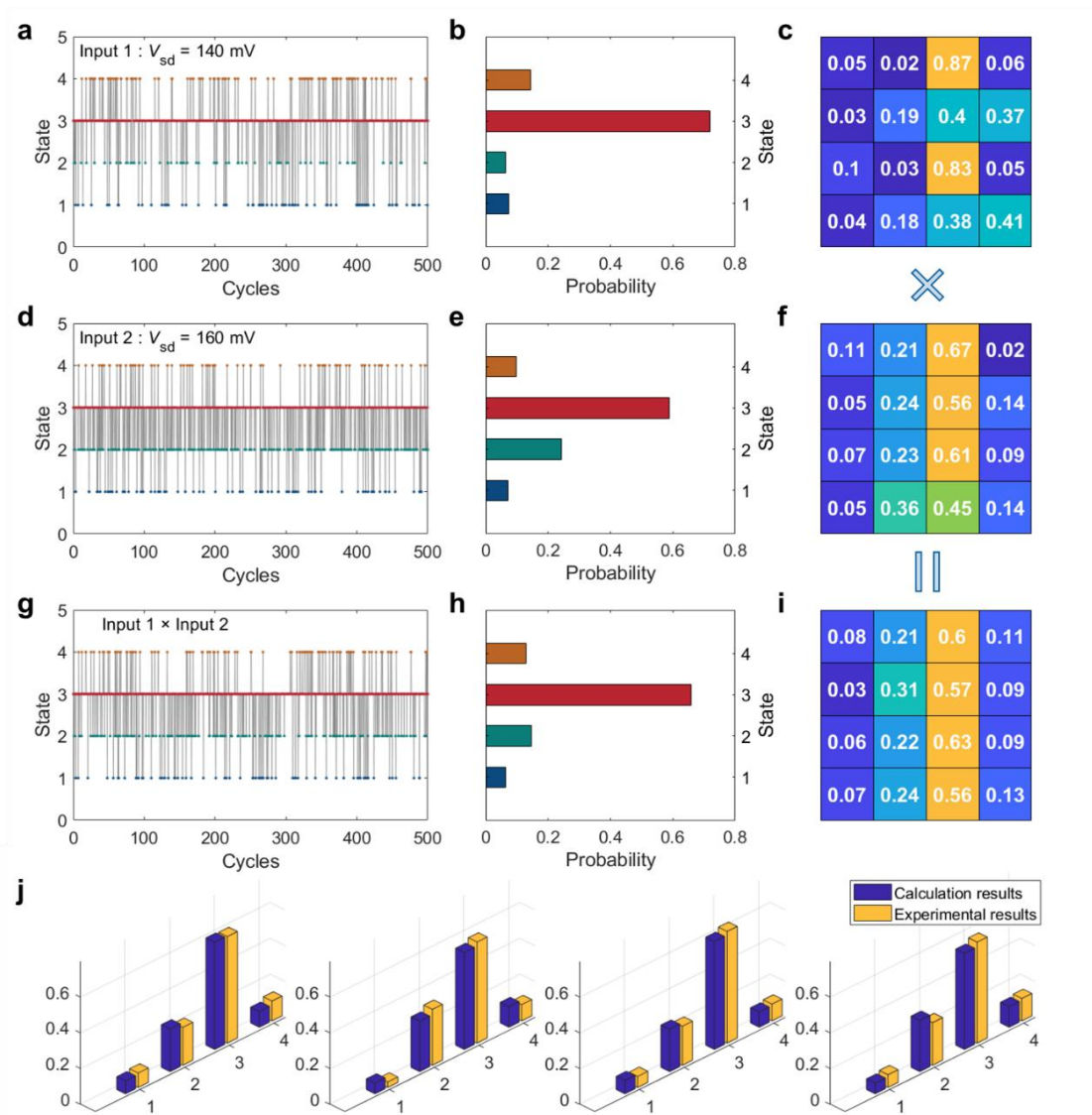

**Fig. 4: High-precision matrix multiplication.**

**a.** State–cycle response curves of Device 2 under $V_{sd}$ = 140 mV. The state of the device at the beginning and end of each cycle was recorded with a period of 1 s and plotted in the curve. Four different conductance states are indicated (a total of 1500 cycles were recorded as shown in **Supplementary information 7**, and only representative sections are shown in the figures). **b.** State distribution histograms derived from the state–cycle responses in a. **c.** State transition matrices derived from the data in a (time interval of 1 s, matrix elements rounded to two decimal places). **d.** State–cycle response curves of

Device 2 under $V_{sd}$ = 160 mV. **e.** State distribution histograms derived from the state–cycle responses in d. **f.** State transition matrices derived from the data in d. **g.** State–cycle response curves of the device obtained by switching $V_{sd}$ between 140 mV and 160 mV. Under a voltage of 140 mV, the initial state of the device is recorded. After a dwell time of 1 s, the voltage is switched to 160 mV, where it remains for another second before the final state of the device is recorded, completing one cycle. **h.** State distribution statistics corresponding to g. **i.** State transition matrix corresponding to g. **j.** Comparison of measured and calculated values of matrix multiplication: yellow data represent the measured values, and blue data represent the values obtained through direct calculation, with the maximum and the average errors less than 0.05 and 0.03, respectively.

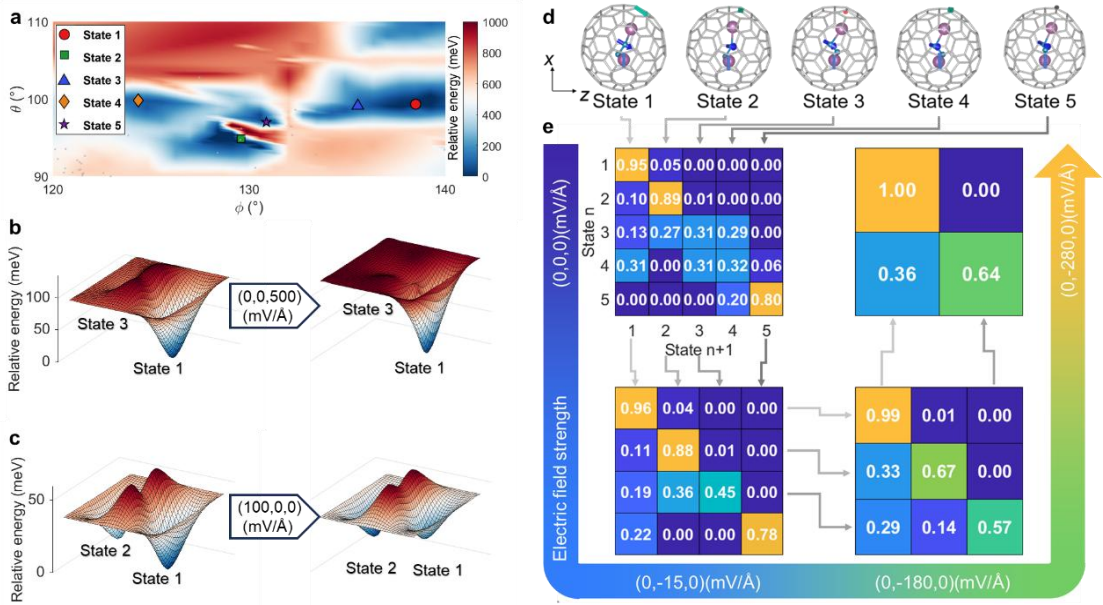

**Fig. 5: Evolution of rich energy landscapes driven by electric fields.**

**a.** PES of the Sc$_2$C$_2$@C$_{88}$ configurations. Different configurations are defined by the orientation of their Sc–Sc bond vector. Here, $\phi$ and $\theta$ correspond to the angles between the Sc–Sc direction and the $z$-axis and $x$-axis, respectively. The red, green, blue, orange and purple points represent the five states with the highest values. All other computed configurations are indicated by black points. **b, c.** Electric field regulation of the PES: (b) barrier modulation; (c) relative energy adjustment. **d.** Geometric configurations and electric dipole moment directions of the lowest energy states. The blue arrows indicate the macroscopic electric dipole moment vectors. Different colored markings on the surface of the carbon cage indicate the adsorption sites for Sc atoms. **e.** Evolution of the interstate transition matrix under different external electric fields.